\documentclass[12pt]{article}
\usepackage{epsf,amsfonts,amssymb,amsbsy}
\def\brtoeq{\beta_{\rm e}}
\begin{document}
\begin{center}
{\bf\large Contributions of leptoquark interactions into the tensor and scalar
form factors of $\boldsymbol{K^+\to \pi^0 l^+ \nu_l}$ decay}\\[3mm]
{\sl  V.V.Kiselev, A.K.Likhoded, V.F.Obraztsov}\\[1mm]
Russian State Research Center ``Institute for High Energy Physics'',\\
Protvino, Moscow Region, 142280 Russia\\
Fax: (0967)-744739, E-mail: kiselev@th1.ihep.su
\end{center}
\begin{abstract}
In the framework of scalar-vector dominance we calculate the hadronic matrix
elements of scalar and tensor effective quark currents induced by virtual
leptoquark interactions. Combined bounds on the product of couplings and
leptoquark masses are obtained from experimental data.
\end{abstract}

\section{Introduction}
Recently, new data on the analysis of $K^+\to \pi^0 l^+ \nu_l$ decays were
published by two collaborations: KEK-E246 \cite{E246} and ISTRA \cite{ISTRA}, 
in addition to the presentation given by the Particle Data Group \cite{PDG}.
So, at present we have got quite precise measurements of characteristics in the
$K_{l3}$ decays, that needs a theoretical interpretation in the framework of
Standard Model (SM) as well as beyond it. Such the study is of interest because
of the experimental search for effects, which can point to the contributions
with the violation of combined CP-parity in the kaon decays, for example, the
transverse T-odd polarization of lepton in $K_{l3\gamma}$ modes \cite{BCL},
that can essentially enrich the information on the CP-breaking dynamics in
addition to the program with the B-mesons \cite{BTeV}. 

The matrix element of decay is parameterized in terms of scalar, vector and
tensor form factors, $f_S$, $f_{\pm}$ and $f_T$, in the following general form
\cite{Chiz}:
\begin{equation}
{\cal M}[K^+\to \pi^0 l^+ \nu_l] = {G_{\rm F} V_{su}}\, 
\bigg[- l_\mu \big(f_+ p^\mu + f_- q^\mu\big) +2 m_K l_S f_S +
{\rm i}\,\frac{f_T}{m_K}\, l_{\mu\nu} p^\mu q^\nu \bigg],
\label{me}
\end{equation}
where the lepton currents are given by the expressions
\begin{eqnarray}
l_\mu & = & \bar\nu_{\rm\scriptscriptstyle L}\gamma_\mu
l_{\rm\scriptscriptstyle L},
\nonumber\\
l_{\mu\nu} & = & \bar\nu_{\rm\scriptscriptstyle L}\sigma_{\mu\nu}
l_{\rm\scriptscriptstyle R},
\nonumber\\
l_S & = & \bar\nu_{\rm\scriptscriptstyle L} l_{\rm\scriptscriptstyle R},
\nonumber
\end{eqnarray}
so that chiral spinors are
$$
\theta_{\rm\scriptscriptstyle R,\,L} = \frac{1}{2}(1\pm\gamma_5)\, \theta,
$$
and $G_{\rm F}$ is the Fermi constant, $m_K$ is the mass of kaon. The
four-momenta are defined as
$$
p = p_K + p_\pi, \quad q = p_K - p_\pi,
$$
while $V_{su}$ is the matrix element of Cabibbo--Kobayashi--Maskawa matrix for
the mixing of weak charged quark-currents. We define the generators $\sigma$ by
the commutator
$$
\sigma_{\mu\nu} =\frac{\rm i}{2}[\gamma_\mu,\gamma_\nu].
$$
The dependence of form factors on $q^2$ is usually expressed in terms of linear
slopes normalized to get the dimensionless quantities
\begin{equation}
\lambda_i = \left.\frac{{\rm d}\ln f_i(q^2)}{{\rm d}
q^2/m_\pi^2}\right|_{q^2=0},
\end{equation}
where $m_\pi$ is the pion mass. The combination of form factors
\begin{equation}
f_0 = f_+ +\frac{q^2}{p\cdot q}\, f_-,
\end{equation}
is introduced, so that the experimental data are given in terms of the
following set \cite{Chiz}:
$$
\lambda_+, \quad \lambda_0, \quad \frac{f_{S,\, T}}{f_+(0)}.
$$
The data of \cite{E246,ISTRA} on $\lambda_{+,0}$ can be averaged, so that with
the statistical errors we get
\begin{eqnarray}
\lambda_+ & = & 0.0287\pm 0.0018,\label{l+}\\
\lambda_0 & = & 0.0203\pm 0.0033,\label{l0}
\end{eqnarray}
while the systematic uncertainties are given in the original papers. The values
in (\ref{l+}) and (\ref{l0}) result in the ratio
\begin{equation}
\frac{f_-(0)}{f_+(0)} = -0.096 \pm 0.043.
\end{equation}
The given parameter $\lambda_+$ is in a good agreement with the PDG values for
both the electron and muon modes \cite{PDG}, while $\lambda_0$ and
${f_-(0)}/{f_+(0)}$ above are within the limits of 1.5\,$\sigma$-deviations
from the PDG averages. The preliminary analysis by KTeV \cite{KTeV} gives
$$
\lambda_+  =  0.0275\pm 0.0008,
$$
which is close to the estimate in (\ref{l+}).

In the framework of SM we get the form factors
\begin{equation}
\langle \pi^0(p_\pi)|\, \bar s \gamma_\mu u\,|K^+(p_K)\rangle =
\frac{1}{\sqrt{2}}\, (f_+ p_\mu + f_- q_\mu) ,
\label{V}
\end{equation}
while 
$$
f_S = f_T = 0.
$$
Therefore, the study of scalar and tensor form factors is a good test for the
search of `new' physics beyond the SM.

Supposing (\ref{V}), we derive
\begin{equation}
q^\mu\, \langle \pi^0(p_\pi)|\, \bar s \gamma_\mu u\,|K^+(p_K)\rangle =
(m_u-m_s)\,\langle \pi^0(p_\pi)|\, \bar s u\,|K^+(p_K)\rangle = (p\cdot q)\,
\frac{f_0}{\sqrt{2}},
\label{divV}
\end{equation}
implying that the form factor $f_0$ determines the matrix element of scalar
quark-current.

The contraction of vector lepton-current
$$
- l_\mu\, q^\mu\, f_- = m_l f_-\, l_S,
$$
induces the scalar term in the matrix element. So, the electron mode is more
sensitive to the extraction of scalar form factor, since the SM background
contribution is suppressed by the lepton mass,
$$
f_S^{\rm SM} = \frac{m_l}{2m_K}\, f_-.
$$
At present, the measurements of $f_S$ and $f_T$ result in values slightly
deviating from zero, that is consistent with the expectations of SM. So, in the
electron mode \cite{E246}
\begin{eqnarray}
\frac{f_S}{f_+(0)} & = & 0.0040\pm 0.0160({\rm stat.})\pm0.0067({\rm syst.}),\\
\frac{f_T}{f_+(0)} & = & -0.019\pm 0.080({\rm stat.})\pm0.038({\rm syst.}),
\end{eqnarray}
where we have taken into account the redefinition of sign in comparison with
the appropriate formula in \cite{E246} as accepted in this paper in (\ref{me}),
while the combined analysis of muon and electron modes in \cite{ISTRA} results
in the similar values
\begin{eqnarray}
\frac{f_S}{f_+(0)} & = & 0.004\pm 0.005({\rm stat.})\pm0.005({\rm syst.}),\\
\frac{f_T}{f_+(0)} & = & -0.021\pm 0.028({\rm stat.})\pm0.014({\rm syst.}).
\end{eqnarray}
The collaboration KTeV presented the following constraints in the electron
mode:
\begin{eqnarray}
\left|\frac{f_S}{f_+(0)}\right| & < & 0.04,\\
\left|\frac{f_T}{f_+(0)}\right| & < & 0.14.
\end{eqnarray}
In the present paper we study nonzero contribution of leptoquark interactions
to the tensor form factor, which correlates with the scalar one due to the
Fierz transformation. In section 2 the effective lagrangians with the virtual
leptoquarks are described as concerns for the decays of $K^+\to \pi^0 l^+
\nu_l$, and the requared matrix elements of quark currents are presented.
General expressions for the hadronic matrix elements with the tensor structure
are derived in section 3, where we develop the model based on the dominance of
vector and scalar mesons and adjust it in the description of $f_{\pm,0}$ form
factors. In the framework of potential approach the preferable region of model
parameter is limited in agreement with the experimental data. The constraints
on the masses of scalar leptoquark and their couplings to the fermions are
obtained in section 4. The results are summarized in the Conclusion.

\section{The contribution of leptoquark interactions}

A consistent classification of leptoquarks under the gauge symmetries of SM
were done by Buch\-m\"uller, R\"uckl and Wyler in \cite{BRW}. We accept the
nomenclature prescribed in \cite{Z} as shown in Table \ref{OPAL} extracted from
\cite{OPAL}. So, the leptoquarks are marked by their spin, representation of
weak $SU(2)$-group (singlets, doublets and triplets), appropriate electric
charges in the multiplets and the fermion number $F$. For the sake of
briefness, the flavor of lepton is marked by the electron in Table \ref{OPAL},
while the couplings ${\rm Y}_{\rm\scriptscriptstyle L,\,R}$ should be labelled
by the flavor indices, too.

\renewcommand{\arraystretch}{1.2}
\begin{table}[htbp]
  \begin{center}
    \begin{tabular}{|c|c|c|cc|c|}\hline   
                                &        &     & \multicolumn{2}{|c|}{coupling
                                and} & \\ 
  scalar LQ ($\tilde{\mbox{q}}$) & charge & $F$ & \multicolumn{2}{|c|}{decay
  mode}   & $\brtoeq$ \\ 
\hline
\hline
\setlength{\unitlength}{1mm}
\raisebox{-3mm}{\begin{picture}(20,10)
\put(0.5,0){\epsfxsize=2cm \epsfbox{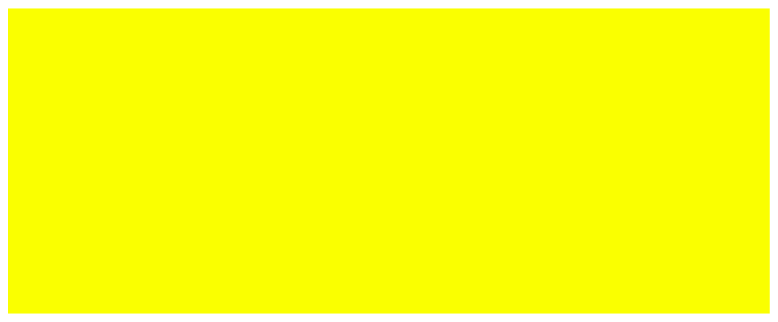}}
\put(0,3){  S$_{\rm 0}$ (or ${\rm \tilde{d}}_{\rm\scriptscriptstyle R}$) }
\end{picture}  
}
  & $-$1/3 & 2 & 
  $\begin{array}{@{}c@{}} {\rm Y}_{\rm\scriptscriptstyle L}: \\ 
  {\rm Y}_{\rm\scriptscriptstyle R}: \end{array}$ &
  \begin{tabular}{@{}c@{}} e$^-_{\rm\scriptscriptstyle L}$u, 
     \raisebox{-1mm}{\begin{picture}(20,10)
\put(0.,0){\epsfxsize=1cm \epsfbox{7.eps}}
\put(1,2){  $\nu_{\rm\scriptscriptstyle L}$d }
\end{picture}  
}
  \\ 
   \raisebox{-1mm}{\begin{picture}(20,10)
\put(0.,0){\epsfxsize=1cm \epsfbox{7.eps}}
\put(2,2){ e$^-_{\rm\scriptscriptstyle R}$u}
\end{picture}  
}
    \end{tabular} &
  \begin{tabular}{@{}c@{}} 1/2 \\ 1 \end{tabular} \\
\hline
  \~S$_{\rm 0}$ & $-$4/3 & 2 & 
  ${\rm Y}_{\rm\scriptscriptstyle R}$: & 
  e$^-_{\rm\scriptscriptstyle R}$d & 
  1 \\ 
\hline
  \~S$_{\rm 1/2}$ (or ${\rm \bar{\rm \tilde{d}}_{\rm\scriptscriptstyle L}}$) &
  +1/3 & 0 &  
  ${\rm Y}_{\rm\scriptscriptstyle L}$: & 
  $\nu_{\rm\scriptscriptstyle L}\bar{\rm d}$ & 
  0 \\
  \~S$_{\rm 1/2}$ (or ${\rm \bar{\rm \tilde{u}}_{\rm\scriptscriptstyle L}}$) &
  $-$2/3 & 0 &  
  ${\rm Y}_{\rm\scriptscriptstyle L}$: & 
  e$^-_{\rm\scriptscriptstyle L}\bar{\rm d}$ & 
  1 \\ 
\hline
  S$_1$ & 
  \begin{tabular}{@{}c@{}} +2/3 \\ $-$1/3 \\ $-$4/3 \end{tabular} &
  2 &
  $\begin{array}{@{}c@{}} {\rm Y}_{\rm\scriptscriptstyle L}: \\ {\rm
  Y}_{\rm\scriptscriptstyle L}: \\ {\rm Y}_{\rm\scriptscriptstyle L}:
  \end{array}$ &
  \begin{tabular}{@{}c@{}} $\nu_{\rm\scriptscriptstyle L}$u \\
                           $\nu_{\rm\scriptscriptstyle L}$d,
                           e$^-_{\rm\scriptscriptstyle L}$u \\
                           e$^-_{\rm\scriptscriptstyle L}$d \end{tabular} &
  \begin{tabular}{@{}c@{}} 0 \\ 1/2 \\ 1 \end{tabular} \\
\hline
\raisebox{-3mm}{\begin{picture}(20,10)
\put(0.,0){\epsfxsize=0.9cm \epsfbox{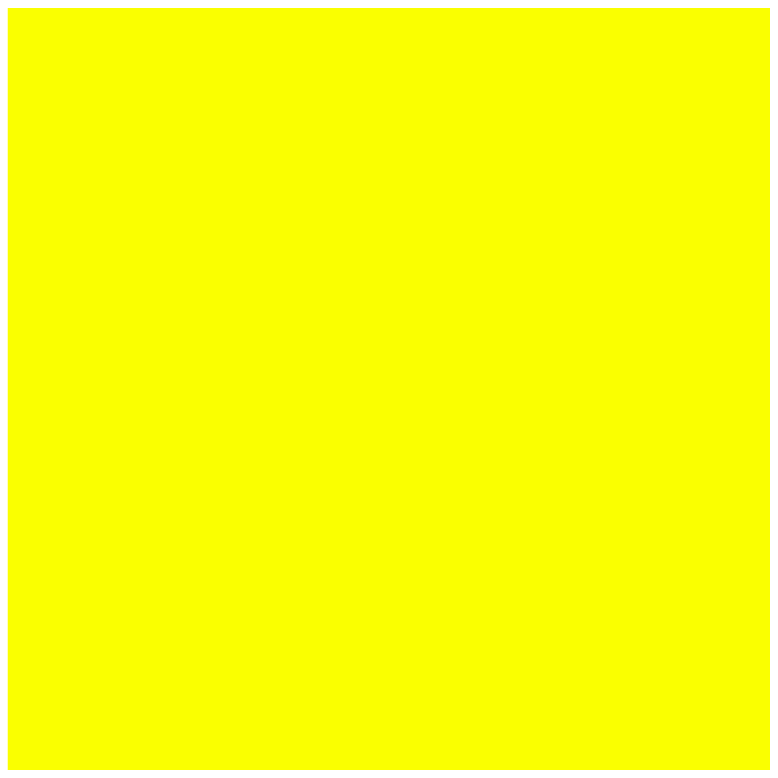}}
\put(0,4){    S$_{1/2}$  }
\end{picture}  
}
  &
  \begin{tabular}{@{}c@{}} $-$2/3 \\ \\ $-$5/3 \end{tabular} &
  0 &
  $\begin{array}{@{}c@{}} {\rm Y}_{\rm\scriptscriptstyle L}: \\ {\rm
  Y}_{\rm\scriptscriptstyle R}: \\ 
                          {\rm Y}_{\rm\scriptscriptstyle L}: \\ {\rm
                          Y}_{\rm\scriptscriptstyle R}: \end{array}$ &
  \begin{tabular}{@{}c@{}} 
  \raisebox{-1mm}{\begin{picture}(20,10)
\put(0.,0){\epsfxsize=1cm \epsfbox{7.eps}}
\put(0,2){   $\nu_{\rm\scriptscriptstyle L}\bar{\rm u}$ }
\end{picture}  
}
  \\
    \raisebox{-1mm}{\begin{picture}(20,10)
\put(0.,0){\epsfxsize=1cm \epsfbox{7.eps}}
\put(0,2){ e$^-_{\rm\scriptscriptstyle R}\bar{\rm d}$ }
\end{picture}  
}
 \\
                           e$^-_{\rm\scriptscriptstyle L}\bar{\rm u}$ \\
                           e$^-_{\rm\scriptscriptstyle R}\bar{\rm u}$
                           \end{tabular} &
  \begin{tabular}{@{}c@{}} 0 \\ 1 \\ 1 \\ 1 \end{tabular} \\
\hline
\multicolumn{6}{c}{}\\
\hline
            &        &     & \multicolumn{2}{|c|}{coupling and} & \\ 
 vector LQ  & charge & $F$ & \multicolumn{2}{|c|}{decay mode}   & $\brtoeq$ \\ 
\hline\hline
  V$_{1/2}$ &
  \begin{tabular}{@{}c@{}} $-$1/3 \\ \\ $-$4/3 \end{tabular} &
  2 &
  $\begin{array}{@{}c@{}} {\rm Y}_{\rm\scriptscriptstyle L}: \\ {\rm
  Y}_{\rm\scriptscriptstyle R}: \\ 
                          {\rm Y}_{\rm\scriptscriptstyle L}: \\ {\rm
                          Y}_{\rm\scriptscriptstyle R}: \end{array}$ &
  \begin{tabular}{@{}c@{}} $\nu_{\rm\scriptscriptstyle L}$d \\
                           e$^-_{\rm\scriptscriptstyle R}$u \\
                           e$^-_{\rm\scriptscriptstyle L}$d \\
                           e$^-_{\rm\scriptscriptstyle R}$d \end{tabular} &
  \begin{tabular}{@{}c@{}} 0 \\ 1 \\ 1 \\ 1 \end{tabular} \\
\hline
  \~V$_{1/2}$ & 
  \begin{tabular}{@{}c@{}} +2/3 \\ $-$1/3 \end{tabular} &
  2 &
  $\begin{array}{@{}c@{}} {\rm Y}_{\rm\scriptscriptstyle L}: \\ {\rm
  Y}_{\rm\scriptscriptstyle L}: \end{array}$ &
  \begin{tabular}{@{}c@{}} $\nu_{\rm\scriptscriptstyle L}$u \\
                           e$^-_{\rm\scriptscriptstyle L}$u \end{tabular} &
  \begin{tabular}{@{}c@{}} 0 \\ 1 \end{tabular} \\
\hline
  V$_0$ & $-$2/3 & 0 &
  $\begin{array}{@{}c@{}} {\rm Y}_{\rm\scriptscriptstyle L}: \\ {\rm
  Y}_{\rm\scriptscriptstyle R}: \end{array}$ &
  \begin{tabular}{@{}c@{}} e$^-_{\rm\scriptscriptstyle L}\bar{\rm d}$,
  $\nu_{\rm\scriptscriptstyle L}\bar{\rm u}$
  \\
                           e$^-_{\rm\scriptscriptstyle R}\bar{\rm d}$
                           \end{tabular} &
  \begin{tabular}{@{}c@{}} 1/2 \\ 1 \end{tabular} \\
\hline
  V$_1$ &
  \begin{tabular}{@{}c@{}} +1/3 \\ $-$2/3 \\ $-$5/3 \end{tabular} &
  0 &
  $\begin{array}{@{}c@{}} {\rm Y}_{\rm\scriptscriptstyle L}: \\ {\rm
  Y}_{\rm\scriptscriptstyle L}: \\ {\rm Y}_{\rm\scriptscriptstyle L}:
  \end{array}$ &
  \begin{tabular}{@{}c@{}} $\nu_{\rm\scriptscriptstyle L}\bar{\rm d}$ \\
                           e$^-_{\rm\scriptscriptstyle L}\bar{\rm d}$,
                           $\nu_{\rm\scriptscriptstyle L}\bar{\rm u}$
                           \\
                           e$^-_{\rm\scriptscriptstyle L}\bar{\rm u}$
                           \end{tabular} &
  \begin{tabular}{@{}c@{}} 0 \\ 1/2 \\ 1 \end{tabular} \\
\hline
  \~V$_0$ & $-$5/3 & 0 & 
  ${\rm Y}_{\rm\scriptscriptstyle R}$: & 
  e$^-_{\rm\scriptscriptstyle R}\bar{\rm u}$ & 
  1 \\ 
\hline
  \end{tabular}
    \caption{The first generation scalar (S) leptoquarks/squarks and vector (V) 
leptoquarks in the BRW model \cite{BRW} according to the nomenclature
in~\cite{Z} with their electric charge in units of $e$ and fermion
number $F=L+3B$.  For each possible non-zero coupling ${\rm Y}$ the
decay modes and the corresponding branching ratio $\brtoeq$ 
for the decay into an electron and a quark are also listed.  The restrictions
on the values of $\brtoeq$ arise from the assumption of chiral couplings.}
    \label{OPAL}
  \end{center}
\end{table}
\renewcommand{\arraystretch}{1.0}

\begin{figure}[th]
\setlength{\unitlength}{1mm}
\begin{center}
\begin{picture}(140,50)
\put(5,1){\epsfxsize=6cm \epsfbox{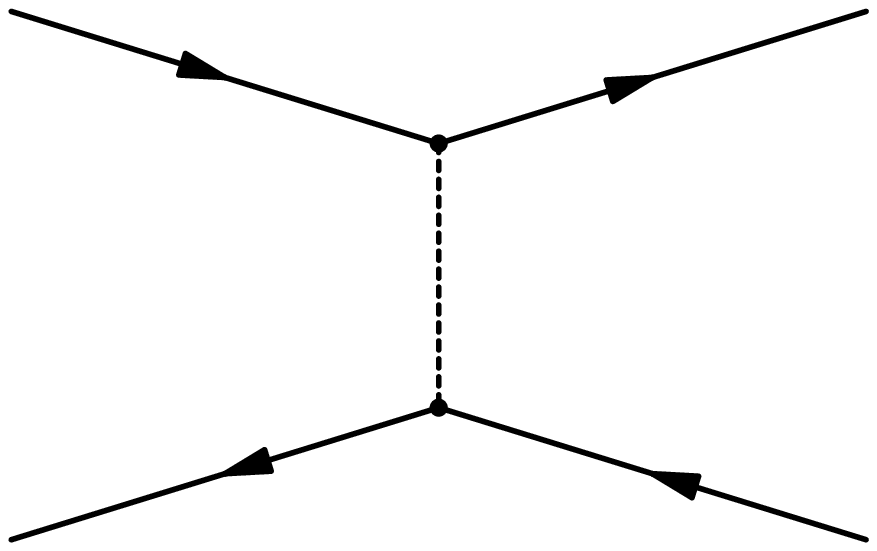}}
\put(75,1){\epsfxsize=6cm \epsfbox{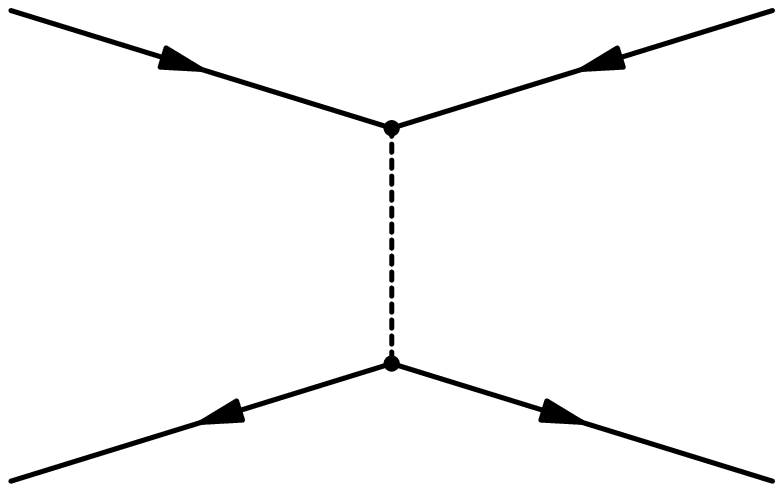}}
%%%%%%%%%%%%%%%%%
\put(3,0){$s$}
\put(3,37){$u$}
\put(65,0){$l_{\rm\scriptscriptstyle\rm R}$}
\put(65,37){$\nu_{\rm\scriptscriptstyle\rm L}$}
\put(36.5,18){$S_{1/2}$}
%%%%%%%%%%%%%%%%%
\put(73,0){$s$}
\put(73,37){$u$}
\put(135,0){$\nu_{\rm\scriptscriptstyle\rm L}$}
\put(135,37){$l_{\rm\scriptscriptstyle\rm R}$}
\put(106.5,18){$S_{0}$}
%%%%%%%%%%%%%%%%%
\end{picture}
\end{center}
\caption{Two kinds of leptoquark exchanges contributing to the tensor form
factor in the decay $K^+\to \pi^0 l^+ \nu_l$. }
\label{diag}
\end{figure}

The diagrams describing the contribution of leptoquark interactions into the
form factors under study are shown in Fig. \ref{diag}.

The tensor terms appear under the Fierz transformations, so that the vector
leptoquarks do not contribute into the tensor form factor. Further, the tensor
term shifts the helicity of leptons. Therefore, we isolate the leptoquarks
involving the interaction with both the left-handed neutrinos and right-handed
charged leptons. The appropriate vertices are shaded in Table \ref{OPAL}.
Thus, we consider the following scalar leptoquarks: the singlet $S_0$ and the
doublet $S_{1/2}$ with the charge $-2/3$.

The Yukawa-like interactions involving the strange quark have the form
\begin{eqnarray}
{\cal L}[S_{1/2}] & = & S^*_{1/2}\, \big({\rm Y}_{\rm\scriptscriptstyle L}\,
\bar u_{\rm\scriptscriptstyle R}\nu_{\rm\scriptscriptstyle L} +
{\rm Y}_{\rm\scriptscriptstyle R}\,
\bar s_{\rm\scriptscriptstyle L} l_{\rm\scriptscriptstyle R}\big) +{\rm
h.c.},\\
{\cal L}[S_{0}] & = & S^*_{0}\, \big[{\rm Y}^{[0]}_{\rm\scriptscriptstyle L}\,
(\bar u_{\rm\scriptscriptstyle C,R}e_{\rm\scriptscriptstyle L}+
\bar s_{\rm\scriptscriptstyle C,R}\nu_{\rm\scriptscriptstyle L})+
{\rm Y}^{[0]}_{\rm\scriptscriptstyle R}\,
\bar u_{\rm\scriptscriptstyle C,L} l_{\rm\scriptscriptstyle R}\big] +{\rm
h.c.},
\end{eqnarray}
where we have omitted the flavor indices. These lagrangians induce the
effective low-energy interactions according to the formulae
\begin{eqnarray}
{\cal L}_{\rm eff} & = & -\frac{1}{8}\, \frac{{\rm Y}_{\rm\scriptscriptstyle R}
{\rm Y}^*_{\rm\scriptscriptstyle L}}{M^2_{S_{1/2}}}\,
(\bar s_{\rm\scriptscriptstyle L}\sigma_{\alpha\beta} u_{\rm\scriptscriptstyle
R})\, (\bar \nu_{\rm\scriptscriptstyle L}\sigma^{\alpha\beta}
l_{\rm\scriptscriptstyle R}) - \nonumber\\
&& -\frac{1}{2}\, \frac{{\rm Y}_{\rm\scriptscriptstyle R}
{\rm Y}^*_{\rm\scriptscriptstyle L}}{M^2_{S_{1/2}}}\,
(\bar s_{\rm\scriptscriptstyle L} u_{\rm\scriptscriptstyle
R})\, (\bar \nu_{\rm\scriptscriptstyle L} l_{\rm\scriptscriptstyle R})+{\rm
h.c.}, \label{eff}
\end{eqnarray}
where we have used the Fierz transformations for the chiral fermions, taking
into account the identity
$$
\gamma_5\, \theta_{\rm\scriptscriptstyle R} = \theta_{\rm\scriptscriptstyle R},
$$
that causes the summation of scalar and pseudoscalar parts (the factor of 2).
The anti-commutation of fermions has been explored, too (the overall negative
sign). Further we introduce the notation
\begin{equation}
\frac{1}{\Lambda^2_{LQ}} = \frac{{\rm Y}_{\rm\scriptscriptstyle R}
{\rm Y}^*_{\rm\scriptscriptstyle L}}{M^2_{S_{1/2}}},
\end{equation}
since the above combination of leptoquark mass and couplings enters the problem
under study.

As for the contribution of $S_0$, one can easily find that the effective
lagrangian has the same form of (\ref{eff}), because the charge conjugation of
spinor is defined by
$$
\theta_{\rm\scriptscriptstyle C} = {\rm C}\, \theta^*,
$$
where ${\rm C} = {\rm i} \gamma_2$ in the Dirac representation of
$\gamma$-matrices, so that
$$
{\rm C} \gamma_\mu^{\rm T} {\rm C}^{-1} = -\gamma_\mu,
$$
where T denotes the transposition. The terms induced by the leptoquarks
$S_{1/2}$ and $S_0$ can interfere, of course. However, we include this effect
into the definition of scale $\Lambda_{LQ}$.

Thus, we can estimate the contribution of leptoquark interactions, once we
calculate the appropriate matrix elements of quark currents, that is the deal
of next section.

\section{Hadronic matrix elements}

\begin{figure}[th]
\setlength{\unitlength}{1.2mm}
\begin{center}
\begin{picture}(80,50)
\put(5,1){\epsfxsize=60\unitlength \epsfbox{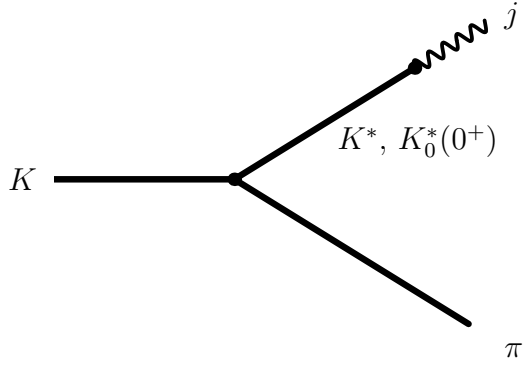}}
%%%%%%%%%%%%%%%%%
\put(0,18.5){$K$}
\put(55,0){$\pi$}
\put(55,37){$j$}
\put(36.5,23){$K^*,\, K^*_0(0^+)$}
\end{picture}
\end{center}
\caption{The diagram describing the contribution of excited vector and
scalar kaon states into the hadronic matrix element of current $j$
factorized from the lepton part in the decay $K^+\to \pi^0 l^+ \nu_l$. }
\label{domin}
\end{figure}

The experimental data on the slopes of form factors shown in the Introduction
are in a good agreement with the estimates in the framework of chiral
perturbation theory ($\chi$PT) \cite{GL}. However, to the moment we have not
any predictions of $\chi$PT on hands as concerns for the hadronic matrix
elements of tensor quark-current. In the present paper we explore the model of
meson dominance, i.e. the dominance of vector and scalar states appropriate for
the quantum numbers of transitions between the quarks. The corresponding
diagram is shown in Fig. \ref{domin}.

Considering the vector quark-current, we can evaluate the form factors
\begin{eqnarray}
f_+(q^2) & = & g_{K^*K\pi}\frac{f_{K^*}\, m_{K^*}}{m^2_{su}(q^2)}\,
\frac{1}{1-q^2/m^2_{su}(q^2)}, \label{fp}\\
f_-(q^2) & = & -f_+(q^2) \frac{m_K^2-m_\pi^2}{m_{su}^2(q^2)}+ g_{K^*_0
K\pi}\frac{f_{K^*_0}}{m_{K^*_0}^2}\,
\frac{1}{1-q^2/m_{K^*_0}^2}, \label{fm}
\end{eqnarray}
in terms of couplings entering the following Lagrangians\footnote{The couplings
$g$ are prescribed for the charged pions, while the neutral ones have the
isospin
factor $1/\sqrt{2}$.}
\begin{eqnarray}
{\cal L}_{K^*K\pi} & = & g_{K^*K\pi}\, (p_K+p_\pi)_\mu\,\epsilon^\mu_{K^*}\;
\varphi_K^*\,\varphi_\pi^*, \\
{\cal L}_{K^*_0 K\pi} & = & g_{K^*_0K\pi}\, \varphi_{K^*_0}\,
\varphi_K^*\,\varphi_\pi^*, 
\end{eqnarray}
and
\begin{eqnarray}
\langle K^*(k)|\, \bar s \gamma_\mu u\, |0\rangle &=&
f_{K^*}\,\epsilon_\mu^{K^*}\, m_{K^*},\\
\langle K^*_0(k)|\, \bar s \gamma_\mu u\, |0\rangle &=& f_{K^*_0}\,k_\mu,
\end{eqnarray}
where $\varphi$ denotes the appropriate field, and $\epsilon$ is the
polarization vector of $K^*$. In (\ref{fp}) we have introduced the running pole
mass $m_{su}(q^2)$ in the transition $s\to u$. The normalization condition is
rather evident
$$
m_{su}(m_{K^*}^2) = m_{K^*},
$$
while we need the value at $q^2=0$, $m_{su} = m_{su}(0)$, since we use the
approximation of linear evolution of form factors,
\begin{eqnarray}
f_+(q^2) & \approx & g_{K^*K\pi}\frac{f_{K^*}\, m_{K^*}}{m^2_{su}}\,
(1+q^2/m^2_{su}), \label{lfp}\\
f_-(q^2) & \approx & -f_+(q^2) \frac{m_K^2-m_\pi^2}{m_{su}^2}+ g_{K^*_0
K\pi}\frac{f_{K^*_0}}{m_{K^*_0}^2}\,
(1+q^2/m_{K^*_0}^2). \label{lfm}
\end{eqnarray}
The evolution of $m_{su}(q^2)$ to $q^2=0$ is expected to be slow in the
framework of model with the meson dominance. We suppose that the spin forces in
the bound state should be suppressed beyond the pole, since they depend on a
density of bound states, which drops outside the poles. So, the spin-averaged
mass of 1S-level in the $\bar s u$ system is known experimentally,
$$
m_{su}[1S] = \frac{1}{4}(m_K+3 m_{K^*}) \approx 793\; {\rm MeV}. 
$$
We expect that
$$
m_{su}[1S] < m_{su}(0) < m_{K^*}.
$$
So, we put
\begin{equation}
m_{su} \approx \frac{1}{2}(m_{su}[1S] + m_{K^*}) \approx 0.85\; {\rm GeV},
\end{equation}
which is inside the systematics uncertainty of the model. Since the
spin-dependent forces are suppressed in the excited P-waves, we put the pole
mass in the scalar sector to be equal to the experimental value of $K^*_0$.

\begin{figure}[th]
\setlength{\unitlength}{1.1mm}
\begin{center}
\begin{picture}(80,30)
\put(5,1){\epsfxsize=60\unitlength \epsfbox{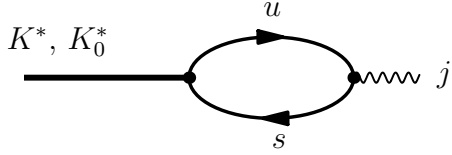}}
%%%%%%%%%%%%%%%%%
\put(55,19){$j$}
\put(3,23){$K^*,\, K^*_0$}
\put(34,27){$u$}
\put(35,11){$s$}
\end{picture}
\end{center}

\vspace*{-14mm}
\caption{The diagram describing the contribution of quark loop into the
hadronic matrix element of current $j$. }
\label{pot}
\end{figure}

From (\ref{lfp}) and (\ref{lfm}) one can easily deduce the expression for the
scalar-channel form factor,
\begin{equation}
f_0(q^2) \approx f_+(0) + q^2\, \frac{g_{K^*_0 K\pi} f_{K^*_0}}
{m_{K^*_0}^2 (m_K^2-m_\pi^2)},
\end{equation}
as well as the slopes,
\begin{eqnarray}
\lambda_+ & = & \frac{m_\pi^2}{m_{su}^2},\label{lp-m}\\
\lambda_0 & = & \delta\cdot \lambda_+,
\end{eqnarray}
where
\begin{eqnarray}
\delta & = & \frac{1}{f_+(0)}\, \frac{g_{K^*_0 K\pi} f_{K^*_0} m_{su}^2}
{m_{K^*_0}^2 (m_K^2-m_\pi^2)},\\[2mm]
f_+(0) & = & g_{K^*K\pi}\frac{f_{K^*}\, m_{K^*}}{m^2_{su}}.
\label{fp-m}
\end{eqnarray}
The most of model parameters can be extracted from the experimental data. So,
the coupling constant $f_{K^*}$ is well known,
$$
f_{K^*} \approx 215\; {\rm MeV},
$$
while the decay constants $g$ are related with the widths measured\footnote{In
the formulae for the total widths of $K^*$ and $K^*_0$ we have explored the
isospin-symmetry relations: $\Gamma[K^{*+} \to K^+\pi^0]=1/2\,\Gamma[K^{*+} \to
K^0\pi^+]$ and the similar equation for the scalar meson $K^*_0$.},
\begin{eqnarray}
\Gamma[K^* \to K\pi] & = & g_{K^*K\pi}^2 \frac{|{\boldsymbol p}_K|^3}{4\pi
m_{K^*}^2},\\
\Gamma[K^*_0 \to K\pi] & = & g_{K^*_0 K\pi}^2 \frac{3\, |{\boldsymbol
p}_K|}{16\pi m_{K^*_0}^2},
\end{eqnarray}
whereas $|{\boldsymbol p}_K|$ denotes the momentum of kaon in the c.m.s, so
that numerically\footnote{In the estimates we put the effective masses in the
equations relating the constants with the total widths, so that $m_{K^*}\to
m_{su}[1S]$ and $m_{K^*_0}\to 2 |{\boldsymbol p}_K|$ in the limit of
$m_{K^*_0}\gg m_{K},m_{\pi}$. In the phenomenological model under study, the
decay constants $g$ enter the form factors in terms of products with the
leptonic couplings $f$. These products should be adjusted in order to satisfy
some conditions motivated by QCD and its chiral symmetry. In this way we have
to follow a specified approach in estimates for both $g$ and $f$ as given
below. We stress that the model parameters $g$ are quite uncertain because of
reasons inherent for the phenomenological approach ignoring higher excitations
as well as a continuum contribution. Neverteless, we argue for the preerable
choice of numerical values.}
$$
g_{K^*K\pi} \approx 3.94,\quad g_{K^*_0 K\pi} \approx 3.48\; {\rm GeV}.
$$
The only free parameter of the model is the coupling $f_{K^*_0}$, which we are
tending to restrict in the framework of potential calculations by the
comparative analysis with the known leptonic constants of $\rho$ and $K^*$. For
this purpose, we calculate the diagram in Fig. \ref{pot}, where the quark-meson
vertex includes the wave function of constituent quarks.

For the scalar state we use the current 
$$
j(x) = \bar s(x) u(x),
$$
with the identity 
$$
{\rm i} \partial_\mu\; [\bar s(x)\gamma^\mu u(x)] = (m_u - m_s) j(x).
$$

In this technique we find
\begin{eqnarray}
f_{K^*_0}^{\rm\scriptscriptstyle PM} & = & \frac{m_s
-m_u}{m_{K^*_0}}\,\frac{18}{m_{\rm red}\,\sqrt{\pi
m_{K^*_0}}}\; |R^\prime_{\bar s u}(0)|,\label{kscalar}\\[2mm]
f_{K^*}^{\rm\scriptscriptstyle PM} & = & \sqrt{\frac{3}
{\pi m_{\bar s u}[1S]}}\; |R_{\bar s u}(0)|,\label{Kstar}
\end{eqnarray}
where $R_{\bar s u}(r)$ denotes the radial wave function in the system of $\bar
s u$, $m_{\bar s u}[1S]$ is the spin-averaged mass of $K^*$ and $K$, and
$m_{\rm red}$ is the constituent reduced mass for $K^*_0$, so that
$$
m_{\rm red} \approx \frac{m_{\bar d u}[1S] m_{\bar s u}[1S]}
{m_{\bar d u}[1S]+m_{\bar s u}[1S]} \approx 0.34\; {\rm
GeV}.
$$
For the $\rho$ meson we have the expression similar to (\ref{Kstar}) under the
substitution $\bar s\to \bar d$.

\begin{figure}[th]
\setlength{\unitlength}{0.8mm}
\begin{center}
\begin{picture}(95,65)
\put(0,1){\epsfxsize=90\unitlength \epsfbox{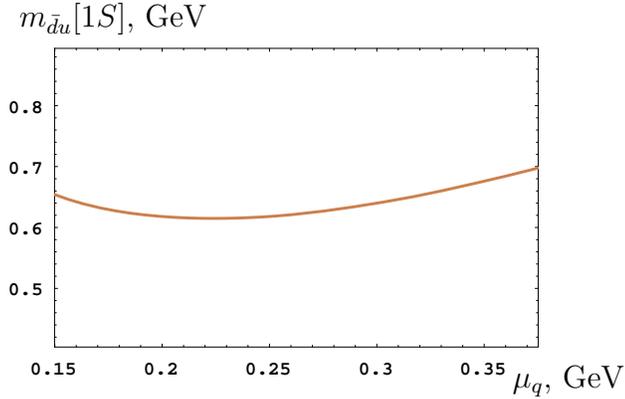}}
%%%%%%%%%%%%%%%%%
\put(3,60){$m_{\bar d u}[1S]$, GeV}
\put(85,0){$\mu_q$, GeV}
\end{picture}
\end{center}

\vspace*{-5mm}
\caption{The mass of $1S$ level in the system $\bar d u$ calculated in the
potential model with the constituent mass $\mu_q$.}
\label{bound}
\end{figure}

Further, we explore the static potential derived in \cite{KKO} and solve the
Schr\"odinger equation 
\begin{equation}
\left[\frac{{\boldsymbol p}^2}{\mu_q} + V(r)\right] \Psi(r) =
[\bar\Lambda(\mu_q)+2(\mu_0-\mu_q)] \Psi(r),
\label{SE}
\end{equation}
for the system $\bar d u$, so that the binding energy $\bar\Lambda(\mu_q)$ of
$1S$-level is related with the mass
\begin{equation}
m_{\bar d u}[1S] =\bar\Lambda(\mu_q)+2 \delta\mu,
\label{m1s}
\end{equation}
and it is shown in Fig. \ref{bound} at $\mu_0 = 0.345$ GeV, $\mu_q^\star
=0.224$ GeV versus the light quark constituent mass $\mu_q$ with $\delta \mu =
\mu_q^\star - \mu_0$. In (\ref{m1s}) we do not add the constituent masses of
light quarks into the mass of meson, since the constituent masses are really
the parts of potential energy $V(r)$ in the confining quark-gluon string.

The mass of bound state shows the minimum versus the constituent mass at
$\mu_q^\star$, which gives the optimal value of mass for the calculation of
radial wave function. For the constituent mass of strange quark we use
$$
\mu_s = m_s + \mu_q^\star,
$$
with $m_s = 0.24$ GeV, which represents the current mass at the scale of 1 GeV
\cite{Narison}.

At this stage the estimates of coupling constants in the potential model can be
optimally got according to (\ref{kscalar}) and (\ref{Kstar}). However, the
corrections by both the quark-gluon loops and a relativistic motion can be
rather essential, that can be taken into account by the introduction of $\cal
K$-factor,
$$
f = {\cal K}\,  f^{\rm\scriptscriptstyle PM}.
$$
At 
$$
{\cal K} = \frac{1}{1.45},
$$
we get the estimates
\begin{eqnarray}
f_\rho &=& 205\; {\rm MeV},\\ 
f_{K^*} &=& 217\; {\rm MeV},\\
f_{K^*_0} &=& 130\; {\rm MeV}. \label{Kscalar}
\end{eqnarray}
The $\cal K$-factor should generally depend on the spin and flavor of current
under srudy. The above estimates show that the dependence on the flavors of
quarks composing the bound state is rather suppressed, since we have amazingly
reproduced the coupling constants of vector states in the limits of
experimental intervals with the uniform $\cal K$-factor. As for the dependence
on the quantum numbers of the meson, we expect that the variation of $\cal
K$-factor is negligibly small because the summed quark spin in both $K^*$ and
$K^*_0$ is equal to 1, while the spin-orbital contributions are usually
suppressed. Thus, the estimate in (\ref{Kscalar}) should be quite accurate up
to 5 MeV, as it does for $\rho$ and $K^*$. Nevertheless, we permit a
conservative variation\footnote{See the sum rule estimates in \cite{Narison}.} 
\begin{equation}
120\; {\rm MeV} < f_{K^*_0} < 140\; {\rm MeV}.
\label{limits}
\end{equation}

\begin{figure}[ph]
\setlength{\unitlength}{1.mm}
\begin{center}
\begin{picture}(95,70)
\put(0,1){\epsfxsize=90\unitlength \epsfbox{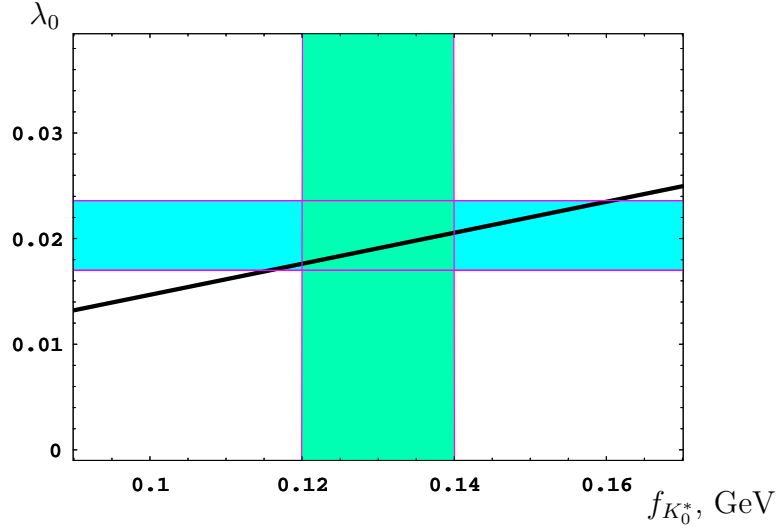}}
%%%%%%%%%%%%%%%%%
\put(3,65){$\lambda_0$}
\put(85,0){$f_{K^*_0}$, GeV}
\end{picture}
\end{center}
\caption{The model predictions for the slope $\lambda_0$ versus the coupling
constant of $K^*_0$ meson (the solid line) in comparison with the
experimental data (the horizontal band). The vertical band gives the region of
preferable values of $f_{K^*_0}$ expected from the potential model. }
\label{f-f0}
\end{figure}

\begin{figure}[th]
\setlength{\unitlength}{1.mm}
\begin{center}
\begin{picture}(95,70)
\put(0,1){\epsfxsize=90\unitlength \epsfbox{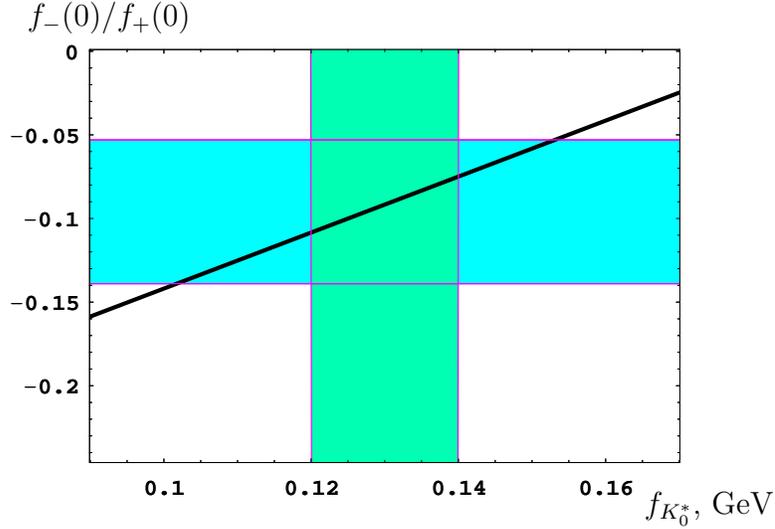}}
%%%%%%%%%%%%%%%%%
\put(3,65){$f_-(0)/f_+(0)$}
\put(85,0){$f_{K^*_0}$, GeV}
\end{picture}
\end{center}
\caption{The model predictions for the ratio $f_-(0)/f_+(0)$ versus the
coupling constant of $K^*_0$ meson (the solid line) in comparison with the
experimental data (the horizontal band). The vertical band gives the region of
preferable values of $f_{K^*_0}$ expected from the potential model. }
\label{f-fpm}
\end{figure}

Further, we can compare the model estimates with the experimental data on
$K_{l3}$ decays listed in the Introduction. This analysis is presented in Figs.
\ref{f-f0} and \ref{f-fpm}. We draw the conclusion on the model is well
adjusted in describing the data. 


According to (\ref{lp-m}) and (\ref{fp-m}) the values of $\lambda_+$ and
$f_+(0)$ are independent of $f_{K^*_0}$. Numerically, we get
\begin{equation}
\lambda_+ =0.0271\pm 0.0011 ,\quad f_+(0) =1.046\pm 0.040,
\end{equation}
which are in a good agreement with both the experimental data and predictions
of $\chi$PT.

Further, we test the model under the Callan--Treiman relation, that expresses
the sum of vector-current form factors in terms of leptonic constants of kaon
and pion:
\begin{equation}
f_+(m_K^2)+f_-(m_K^2) = \frac{f_K}{f_\pi}.
\label{CT}
\end{equation}
In the model under study we get
\begin{equation}
f_+(m_K^2)+f_-(m_K^2) = f_+(0) + \frac{g_{K_0^* K \pi} f_{K_0^*}}{m_{K_0^*}^2 -
m_K^2} \approx f_+(0) + \frac{g_{K_0^* K \pi} f_{K_0^*}}{m_{K_0^*}^2},
\label{CTmd}
\end{equation}
where we have neglected the kaon mass with respect to the scalar meson one.
Then, numerically the relations result in
$$
\frac{f_K}{f_\pi} = 1.305\pm 0.020\quad {\rm or}\quad 
\frac{f_K}{f_\pi} \approx 1.273\pm 0.017
$$
under the variation in (\ref{limits}). So, at $f_\pi=132$ MeV we deduce
$$
f_K = 172\pm 3\;{\rm MeV} \quad {\rm or}\quad 
f_K = 168\pm 4\;{\rm MeV}.
$$
The systematic error caused by the approximation, as we see, is about 5 MeV,
and conservatively one expects
$$
f_K = 170\pm 4\pm 5\;{\rm MeV},
$$
which is in a good agreement with the known data.

Neglecting both the deviation of $f_+(0)$ from the unit and the kaon mass with
respect to the mass of scalar $K_0^*$, we can derive from (\ref{CT}) and
(\ref{CTmd}) the Dashen--Weinstein relation
$$
\lambda_0 = \frac{m_\pi^2}{m_K^2-m_\pi^2}\, \left(\frac{f_K}{f_\pi}-1\right).
$$
Both relations by Callan--Treiman and Dashen--Weinstein can acquire valuable
numerical corrections in the model under study as well as in the $\chi$PT. So,
from the formula for $\lambda_0$ we get
$$
f_K = 160\pm 4\;{\rm MeV},
$$
so that the displacement of $f_K$ points to the possible size of corrections.

Then, we calculate the expression for the hadronic matrix element of tensor
quark-current
\begin{equation}
\langle \pi^0(p_\pi)|\, \bar s \sigma_{\mu\nu} u\,|K^+(p_K)\rangle = -{\rm i}\,
\frac{f_+(q^2)}{\sqrt{2}\,m_{K^*}} (p_\mu q_\nu - p_\nu q_\mu)\approx 
-{\rm i}\, \frac{f_+(0)}{\sqrt{2}\,m_{K^*}} (p_\mu q_\nu - p_\nu q_\mu),
\label{sigmaH}
\end{equation}
where we have neglected the dependence of $f_+$ on $q^2$, since the
antisymmetric tensor is linear in $q$. Formula (\ref{sigmaH}) can be compared
with the general expression 
\begin{equation}
\langle \pi^0(p_\pi)|\, \bar s \sigma_{\mu\nu} u\,|K^+(p_K)\rangle = -
\frac{\rm i}{\sqrt{2}}\,
{\cal B}\, (p_\mu q_\nu - p_\nu q_\mu),
\label{sigmaB}
\end{equation}

\noindent
so that $\cal B$ depends on a single additional quantity $c_-(q^2)$
$$
{\cal B} = c_-\,\frac{f_0(q^2)}{m_s-m_u} + (m_s+m_u)\, \frac{f_-(q^2)}
{p\cdot q},
$$
where we have explored the definition
$$
\langle \pi^0(p_\pi)|\, {\rm i}\bigg\{\bar s (\partial_{\mu} u) - (\partial_\mu
\bar s) u \bigg\}\,|K^+(p_K)\rangle = (c_+\, p_\mu +c_-\,q_\mu)\; 
\langle \pi^0(p_\pi)|\, \bar s  u\,|K^+(p_K)\rangle ,
$$
with an evident condition of self-consistency
$$
c_+ = -\frac{1}{p\cdot q}\, (m_s^2-m_u^2 + c_-\, q^2).
$$
In the model of meson dominance we get
\begin{equation}
c_- = \frac{f_+}{f_0}\,\frac{m_s-m_u}{m_{K^*}} - \frac{f_-}{f_0}\,
\frac{m_s^2-m_u^2}{p\cdot q}.
\end{equation}
Neglecting the current mass of light quark, at $q^2=0$ we find
\begin{eqnarray}
c_-(0) & \approx & \frac{m_s}{m_{K^*}} +(\lambda_+-\lambda_0)\,
\frac{m_s^2}{m_\pi^2},\\
c_+(0) & \approx & -\frac{m_s^2}{m_K^2- m_\pi^2}.
\end{eqnarray}
The physical meaning of $c_\pm$ is rather simple: they determine the difference
between the fractions of meson momenta carried by the $\bar s$ and $u$ quarks
in the kaon and pion under the weak transition. At $q^2 = 0$ we get
\begin{eqnarray}
\alpha_{K} & = & \frac{1}{2}(c_+ + c_-)\approx 0.018,\\
\alpha_{\pi} & = & \frac{1}{2}(c_- - c_+)\approx 0.28.
\end{eqnarray}

\subsubsection*{Spin effects: the polarization of lepton in SM}
Neglecting the suppresed contributions by the scalar and tensor form factors in
(\ref{me}), we calculate the matrix element squared for the lepton with the
spin polarization $s_\alpha$, satisfying the conditions of $s^2 =-1$ and
$s\cdot p_l=0$. Then, omiting an irrelevant overall normalization factor and
putting the vector form factors to be real, we get
\begin{eqnarray}
|{\cal M}[K^+\to \pi^0 l^+ \nu_l]|^2 & \sim &
2 f_+^2\, (p_l \cdot  p)\, (p_\nu \cdot  p) 
+2\,{f_+}\, {f_-}\, m_l^2\, (p_\nu \cdot  p)
+ (p_l \cdot  p_\nu)\,
\big(f_-^2\, m_l^2 - f_{+}^{2}\, {p^2}\big) -
\nonumber \\[2mm] &&
2 f_{+} f_{-}\,m_l\,(p_l \cdot p_\nu)\, (p \cdot s) 
-2f_+^2\,m_l (p_\nu\cdot p)\, (p \cdot  s) +
\label{polme2}
\\[2mm] &&
 2{f_+}{f_-}\,{m_l}\, (p_l \cdot p)\,(p_\nu \cdot s) 
+ m_l\, \big({p^2}\, f_{+}^{2}+f_{-}^{2}\, m_{l}^{2}\big)\,
(p_\nu \cdot  s).
\nonumber 
\end{eqnarray}
Following the ordinary definition for the polarization $P_i$,
$$
|{\cal M}|^2 = \rho_0\left(1-\sum_{i=L,N,T}P_i\, (e_i\cdot s_i)\right),
$$
where $e_L$ denotes the longitudinal four-vector
$$
e_L = \frac{1}{m_l|\boldsymbol p_l|}\,(|\boldsymbol p_l|^2,
E_l \boldsymbol p_l),
$$
the unit vector $e_N$ lies in the decay plane, and it is orthogonal to $e_L$,
while $e_T$ is transversal to both $e_L$ and $e_N$. 

\begin{figure}[ph]
\setlength{\unitlength}{0.7mm}
\begin{center}
\begin{picture}(100,100)
\put(5,5){\epsfxsize=90\unitlength \epsfbox{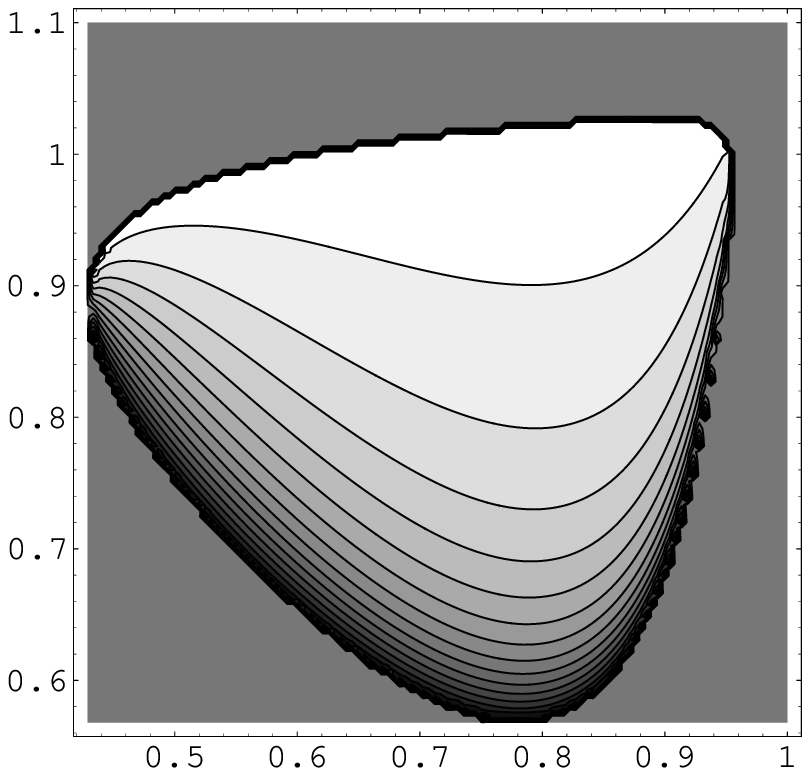}}
\put(85,0){$y=2 E_\mu/m_K$}
\put(3,98){$x=2 E_\pi/m_K$}
\put(59,63){$\scriptscriptstyle -0.94$}
\put(59,47){$\scriptscriptstyle -0.81$}
\put(59,38){$\scriptscriptstyle -0.69$}
\put(59,32){$\scriptscriptstyle -0.56$}
\end{picture}
\begin{picture}(100,95)
\put(5,25){\epsfxsize=90\unitlength \epsfbox{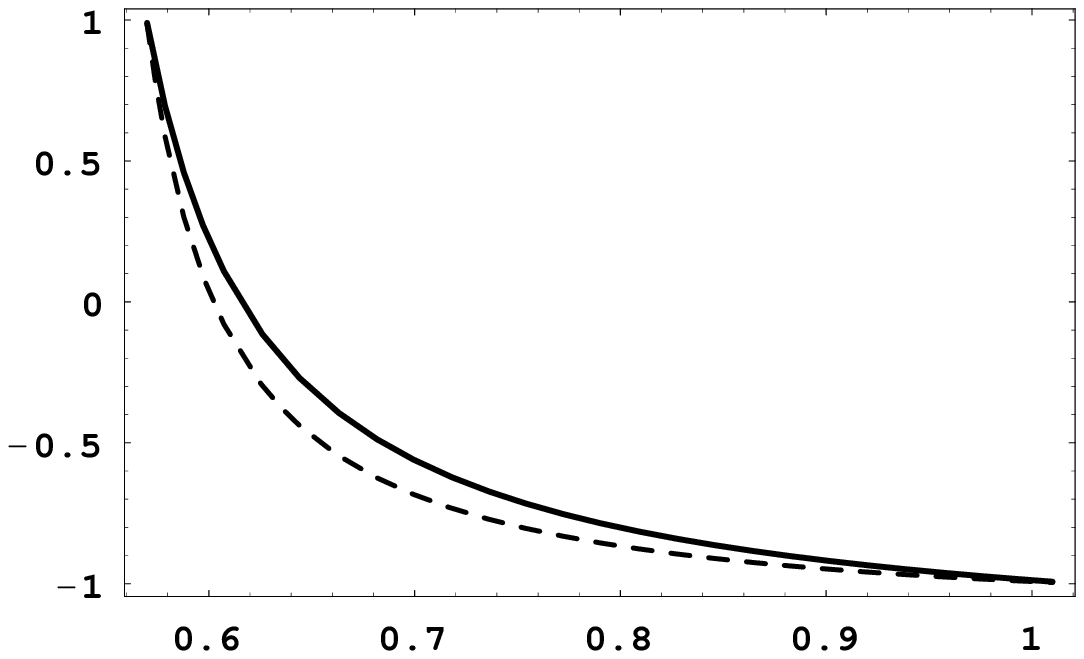}}
\put(85,20){$x$}
\put(55,50){$y=0.8$}
\put(3,82){$P_L$}
\end{picture}
\end{center}
\caption{The distribution of longitudinal polarization on the Dalitz plot for
the decay of $K^+\to \pi^0 \mu^+ \nu_\mu$ with $x=2 E_\pi/m_K$, $y=2
E_\mu/m_K$ (the left picture). The contours are shown with the arithmetic
progression for the polarization: $-0.9375+j\cdot 0.125$. The section of Dalitz
plot at $y=0.8$ with the model value of $f_-$ described in the text (the solid
curve) in comparison with the fixed $f_-=-0.5$ result (the dashed curve).}
\label{mu}
\end{figure}
\begin{figure}[ph]
\setlength{\unitlength}{1mm}
\begin{center}
\begin{picture}(160,70)
\put(0,5){\epsfxsize=7cm \epsfbox{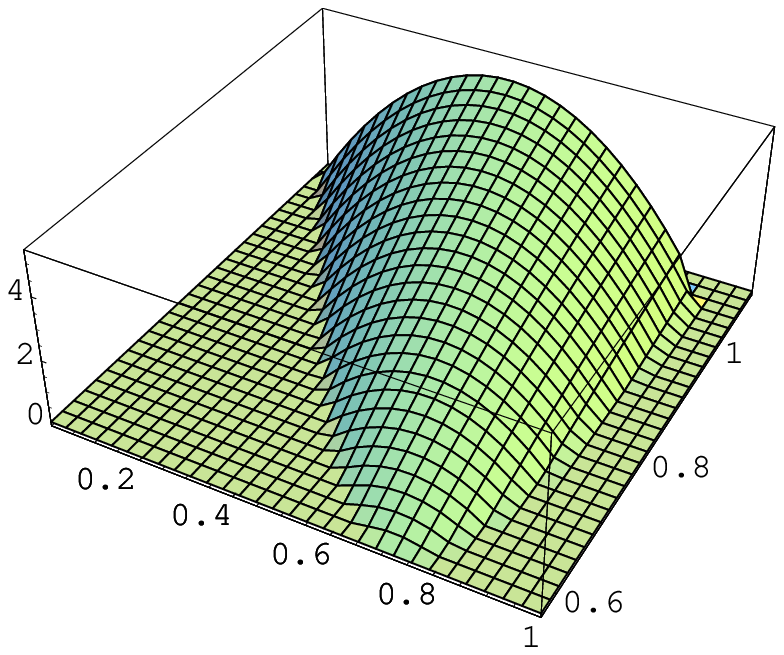}}
\put(80,5){\epsfxsize=7cm \epsfbox{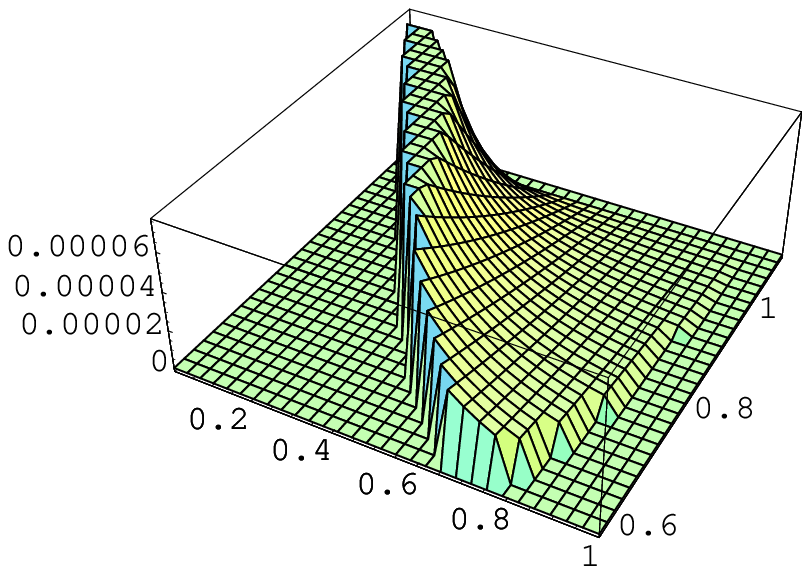}}
\put(0,43){$\rho_0$}
\put(25,10){$y$}
\put(66,18){$x$}
\put(85,43){$\rho_{\Uparrow}$}
\put(105,10){$y$}
\put(146,18){$x$}
\end{picture}
\end{center}
\caption{The Dalitz plots for the decay of $K^+\to \pi^0 e^+ \nu_e$ with the
square of matrix element summed over the electron polarizations (the left
picture, $\rho_0$ in arbitrary units) and with the spin polarization along the
electron momentum (the right picture, $\rho_\Uparrow$ in the same units as
$\rho_0$).}
\label{el}
\end{figure}

The scalar products in (\ref{polme2}) can be easily expressed in terms of pion
and lepton energies. So, for the longitudinal polarization we get
$$
\begin{array}{rcl}
p_\nu\cdot s & = & 
\displaystyle
\frac{1}{m_l|\boldsymbol p_l|}\, [E_\nu |\boldsymbol p_l| -
E_l^2 E_\nu + E_l\, (p_l\cdot p_\nu)], \\[6mm]
p\cdot s & = & 
\displaystyle
\frac{1}{m_l|\boldsymbol p_l|}\, [E_\pi |\boldsymbol p_l| -
E_l^2 E_\pi + E_l\, (p_l\cdot p_\pi)]+\frac{m_K |\boldsymbol p_l|}{m_l}.
\end{array}
$$
The results for the muon and electron modes are shown in Figs. \ref{mu} and
\ref{el}, respectively.

As we can expect from (\ref{polme2}), the variation of longitudinal
polarization is significant for the muon in contrast to the electron, since the
spin dependence exhibits the suppression by the lepton mass. The longitudinal
polarization effects for the electron mode are negligibly small. 

In Fig. \ref{mu} we see that the longitudinal polarization of the muon is
essentially negative in the dominant part of Dalitz plot. The significant
variation of form factor $f_-$ by a factor of 5 shows a weak sensetivity of
longitudinal polarization to $f_-$. Fig. \ref{el} exhibits that the deviation
of electron polarization from $-1$ is negligible.

\begin{figure}[th]
\setlength{\unitlength}{0.7mm}
\begin{center}
\begin{picture}(100,100)
\put(5,5){\epsfxsize=90\unitlength \epsfbox{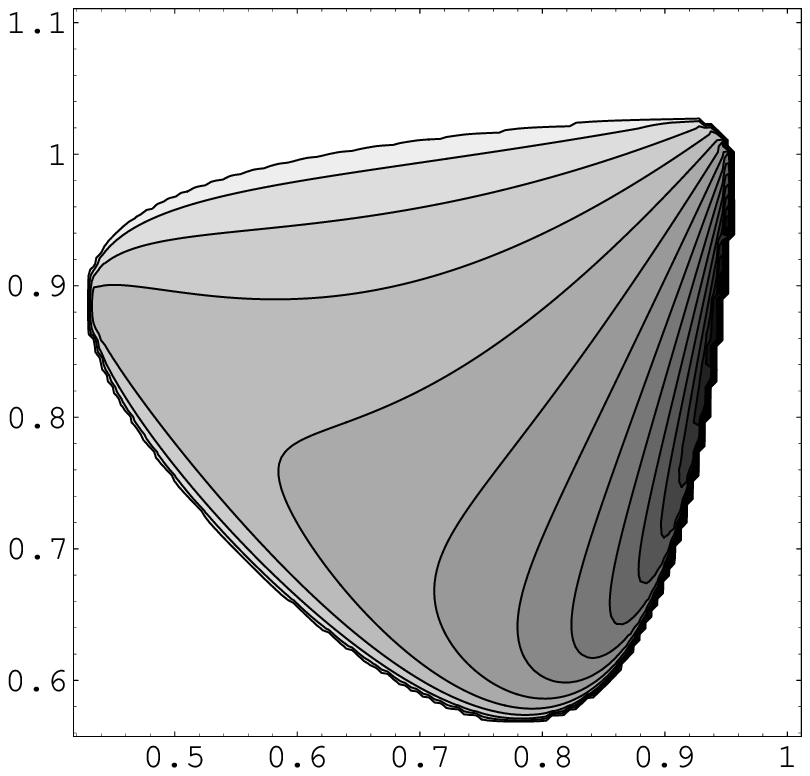}}
\put(85,0){$y$}
\put(3,98){$x$}
\put(33,98){$P_N[\mu]$}
\put(33,61){$\scriptscriptstyle -0.21$}
\put(34,47){$\scriptscriptstyle -0.26$}
\put(44,32){$\scriptscriptstyle -0.32$}
\end{picture}
\begin{picture}(100,100)
\put(5,5){\epsfxsize=90\unitlength \epsfbox{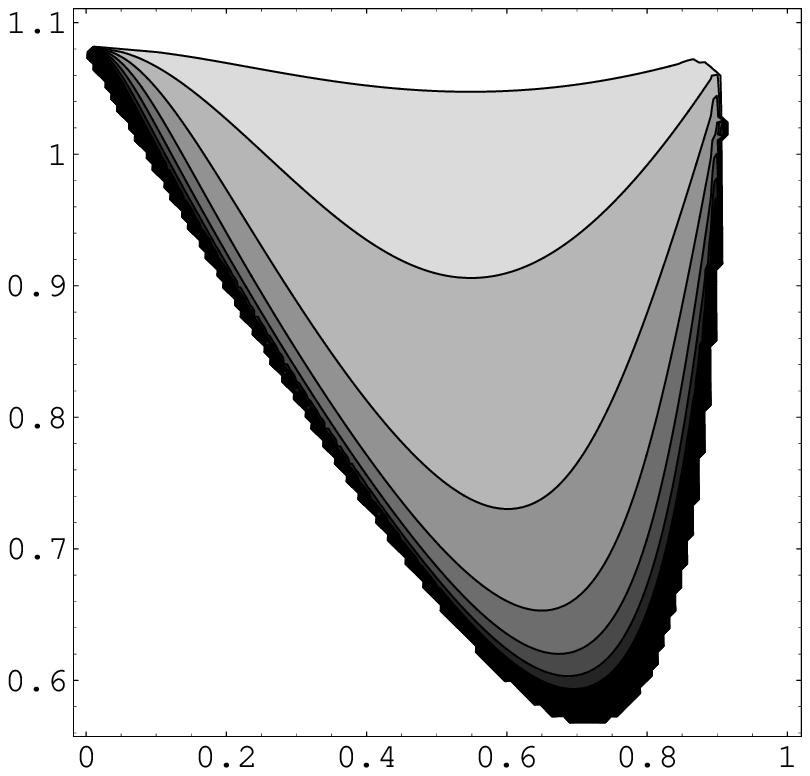}}
\put(85,0){$y$}
\put(3,98){$x$}
\put(33,98){$P_N[e]\cdot 10^2$}
\put(53,64){$\scriptscriptstyle -0.11$}
\put(57,39){$\scriptscriptstyle -0.17$}
\put(60,27.5){$\scriptscriptstyle -0.23$}
\end{picture}
\end{center}
\caption{The distribution of normal polarization on the Dalitz plot for
the decays of $K^+\to \pi^0 \mu^+ \nu_\mu$ (the left picture) and $K^+\to \pi^0
e^+ \nu_e$ (the right picture). The contours are shown with the arithmetic
regressions for the polarization: the muon $-0.0525-j\cdot 0.0525$, the
positron $10^{-3}\cdot(-0.45-j\cdot 0.60)$.}
\label{muN}
\end{figure}

Further, we have calculated the normal polarization of leptons as shown in
Fig. \ref{muN}. We see that the muon has a significant variation
of normal polarization, while the electron spin in the normal direction is
quite small (about 0.2\%), and it is sizable (about 10\%) in the region, where
the number of events is essentailly suppressed.

\section{Constraints on the leptoquark scales}
Under the determination of hadronic matrix elements of quark currents we derive
the ratios of form factors due to the contribution of leptoquark interactions,
\begin{eqnarray}
\frac{f_S}{f_+(0)} & = & \frac{\sqrt{2}}{16 G_{\rm F}
|V_{su}|}\,\frac{m_K^2-m_\pi^2}{(m_s-m_u) m_{K}}\,
\frac{1}{\Lambda_{LQ}^2},\\[2mm]
\frac{f_T}{f_+(0)} & = & -\frac{\sqrt{2}}{32 G_{\rm F}
|V_{su}|}\,\frac{m_K}{m_{K^*}}\, \frac{1}{\Lambda_{LQ}^2},
\end{eqnarray}
where we have supposed the positive definiteness of Yukawa-constant products
with respect to the mixing $V_{su}$. Then we extract the values of leptoquark
scales in the tensor part,
\begin{equation}
\Lambda_{LQ} = 0.48_{-0.17}^{+\infty}\; {\rm TeV},
\end{equation}
while the scalar form factor gives more stringent limit
\begin{equation}
\Lambda_{LQ} = 3.4_{-1.1}^{+\infty}\; {\rm TeV}.
\end{equation}
Thus, we deduce the 95\%-confidence level
$$
\Lambda_{LQ} > 1.2\; {\rm TeV}.
$$
Let us compare the above restriction on the parameters of leptoquark 
interactions with the constraints following from other processes relevant to 
the effective vertices induced by diagrams in Fig. \ref{diag}. Since the 
Yukawa constants are flavor dependent, the direct constraints can be 
obtained from the leptonic decays of kaon, \textit{viz.}, from both the 
electron and muon ones. In this way, the tensor interaction does not 
contribute, while the scalar one results in the multiplicative scaling of 
the decay amplitude. The factor has the form
\begin{equation}
\mathcal{K}_{LQ} \approx 1 - 2\, \frac{f_S}{f_+(0)}\, \frac{m_K}{m_l},  
\end{equation}
where we have neglected the masses of pion and $u$-quark, and $m_l$ denotes 
the mass of lepton. 

The leptonic modes are measured with the accuracy of branching ratios
$$
\frac{\delta \cal B_{\rm e}}{\cal B_{\rm e}} \approx \frac{1}{22},\quad 
\frac{\delta \cal B_{\mu}}{\cal B_{\mu}} \approx \frac{1}{300},
$$
that can be used in order to restrict the scalar interactions induced by the 
leptoquarks. So, taking the ratio of branching ratios, which is independent of
both the leptonic constant of kaon and the CKM element $|V_{su}|$, we get the
expression
$$
\frac{\cal B_{\rm e}}{\cal B_{\mu}} = \frac{m_e^2}{m_\mu^2}\,
\frac{\displaystyle  1 - 4\,
\frac{f_S}{f_+(0)}\, \frac{m_K}{m_e}}{\displaystyle 1 - 4\,
\frac{f_S}{f_+(0)}\, \frac{m_K}{m_\mu}} = 2.3372\cdot 10^{-5}\cdot 
\frac{\displaystyle 1 - 4\, \frac{f_S}{f_+(0)}\,
\frac{m_K}{m_e}}{\displaystyle 1 - 4\, \frac{f_S}{f_+(0)}\, \frac{m_K}{m_\mu}},
$$
where we expand in small corrections following from the leptoquark
interactions. Comparing with the experimental result
$$
\left.\frac{\cal B_{\rm e}}{\cal B_{\mu}}\right|_{\rm exp.} =(2.44\pm
0.11)\cdot 10^{-5},
$$
we find\footnote{As for some other restrictions see ref. \cite{Gershtein:gp},
where the bounds are very similar to those of obtained in the present paper.}
$$
\Lambda_{LQ} > 43\; {\rm TeV}.
$$
Thus, the measurements of semileptonic kaon decay provide us with the 
soft confirmation of constraints following from the leptonic decays, since 
the tensor and scalar effective vertices correlate in the leptoquark 
interactions.

\section{Discussion}
In this paper we have developed a model of meson dominance, which has allowed
us to get quite an accurate description of hadronic form factors in the decay
$K^+\to \pi^0 l^+ \nu_l$. In this way we have adjusted the model under the
experimental data on the matrix element of vector quark-current and calculated
the matrix element of tensor current induced by the leptoquark interactions.
The experimental data on the semileptonic decay of kaon allow us to extract the
constraints on the contributions beyond the Standard Model, so that
$$
\Lambda_{LQ} > 1.2\; {\rm TeV},
$$
where $\Lambda_{LQ}$ represents the ratio of leptoquark mass to the square of
Yukawa-like coupling. This limit softly confirms the bounds following from the
leptonic decays of kaon.

The authors thank the DELPHI and LHCb collaborations for a kind hospitality
during their visit to CERN, where this paper was in progress.

This work was in part supported by the Russian foundation for basic research,
grants 01-02-99315, 01-02-16585 and 00-15-96645, the Russian Ministry on the
education, grant E00-3.3-62.

\end{document}